\documentclass[aps,preprintnumbers,twocolumn,amsmath,amssymb,floatfix,pra]{revtex4-1}
\pdfoutput=1
\usepackage[utf8]{inputenc}
\usepackage{amsmath}
\usepackage{amsfonts}
\usepackage{braket}
\usepackage{tensor}
\usepackage{bbold}
\usepackage{graphicx}
\usepackage{tensor}
\usepackage[colorlinks=true,linkcolor=blue, citecolor=blue, urlcolor=blue, bookmarks]{hyperref}

\def\be{\begin{equation}}
\def\ee{\end{equation}}
\def\bea{\begin{eqnarray}}
\def\eea{\end{eqnarray}}

\def\Xint#1{\mathchoice
       {\XXint\displaystyle\textstyle{#1}}%
       {\XXint\textstyle\scriptstyle{#1}}%
       {\XXint\scriptstyle\scriptscriptstyle{#1}}%
       {\XXint\scriptscriptstyle\scriptscriptstyle{#1}}%
       \!\int}
\def\XXint#1#2#3{{\setbox0=\hbox{$#1{#2#3}{\int}$}
         \vcenter{\hbox{$#2#3$}}\kern-.5\wd0}}
\def\ddashint{\Xint=}
\def\dashint{\Xint-}

\begin{document}

\title{Local correlations in the attractive 1D Bose gas: from Bethe ansatz to the Gross-Pitaevskii equation}
\author{Lorenzo Piroli}
\author{Pasquale Calabrese}
\affiliation{SISSA and INFN, via Bonomea 265, 34136 Trieste, Italy. }

\begin{abstract}
We consider the ground-state properties of an extended one-dimensional Bose gas with pointwise attractive interactions. 
We take the limit where the interaction strength goes to zero as the system size increases at fixed particle density. 
In this limit the gas exhibits a quantum phase transition. 
We compute local correlation functions at zero temperature, both at finite and infinite size. 
We provide analytic formulas for the experimentally relevant one-point functions $g_2$, $g_3$ and analyze their finite-size corrections.
Our results are compared to the mean-field approach based on the Gross-Pitaevskii equation which yields the exact results 
in the infinite system size limit, but not for finite systems.
\end{abstract}

\maketitle

\section{Introduction}\label{intro}
The study of one-dimensional quantum integrable models has produced many remarkable results over the past fifty years. Among its greatest successes, is the derivation of thermodynamical properties of extended systems from the underlying microscopic quantum theory. A unified theoretical framework of integrability is now well established as reported in many excellent textbooks \cite{baxter, gaudin, korepin, takahashi,sutherland,essler}. Until recently, however, the interest for these  studies has been mainly academic, due to the lack of experimental applications.

The situation has completely changed during the past two decades, due to the new experimental possibilities coming from the physics of ultra-cold atoms. Indeed, optical and magnetic traps can nowadays be employed to effectively confine atoms in one spatial dimension where almost ideal Hamiltonians are engineered with a high degree of isolation and control over the experimental parameters \cite{bdz-08,ccgo-11}. Thus the results of exact calculations in integrable models can be tested in cold atomic laboratories, offering a playground where theory and experiments can be compared directly and without ambiguity. 

One of the prototypical examples of integrable models is the one-dimensional Lieb-Liniger gas, describing a system of bosons with pointwise interactions. This model has a long history  \cite{ll-63, mcguire-64,yy-69} and has been intensively studied in the literature, but the exact computation of correlation functions still represents a remarkable theoretical challenge. At the same time, this problem is of paramount importance for a comparison with cold atomic realizations of confined bosons, where quantum correlations are routinely measured in experiments \cite{ksfb-02,spth-02, kww-04,pwmm-04,tohp-04,kww-05,kww-06, awkd-08, jabk-11,fcff-11, hgmd-09,fp-15,mpml-15}. 

In the case of repulsive interactions, a significant amount of theoretical work has already been devoted to the computation of correlation functions \cite{jm-81, slavnov-90, kks-97, gs-03,kgds-03, ffgw-03, od-03, cc-06, ffm-06, ccs-07,ig-08, kmt-09,pgb-09, kms-11, kt-11, kci-11, spci-12,pozsgay-11, pc-14,pk-13, kp-14, kozlowski-14, pc-15, zwkg-16, nrtg-16}. Over the years this problem has inspired the development and application of sophisticated techniques based, for example, on the Bethe ansatz method \cite{ slavnov-90, kks-97, cc-06, kms-11, kt-11, pozsgay-11, kozlowski-14, pc-15} or on field theoretical approaches \cite{kmt-09, kci-11}. It is worth mentioning that while the focus has been traditionally on ground states and thermal states, the past few years have also witnessed an increasing interest in the computation of correlation functions in arbitrary excited states of integrable systems \cite{pozsgay2-11,pozsgay-13,mp-14,ga-15_1,psg-15,bpc-16,pozsgay-16, dp-15}, also in connection with its relevance in the study of non-equilibrium dynamics of one-dimensional Bose gases \cite{ grd-10,ia-12,mc-12,csc-13, dwbc-14, mckc-14,ga-14,zwkg-15,ga-15, dpc-15, dc-14,kcc-14,bck-15,vwed-16, pce-16,pce2-16}.

Attractive interactions have been less studied in the literature \cite{mcguire-64,ccr-00_II, kavoulakis-03,ksu-03,ma-05,ssac-05,sdd-07,ol-07,cc-07,kcu-09,kcu-10,hoc-14,ffp-16}. In this case, the traditional thermodynamic limit of the model 
is ill-defined, with divergences in the ground state energy and in local correlation functions \cite{takahashi, mcguire-64}. These divergences reflect the physical property that strong attractive interactions lead to instabilities in a gas containing a large number of bosons. A stable, non-thermal stationary state can nevertheless be obtained in the thermodynamic limit as a result of an interaction quench, as for the super Tonks- Girardeau gas \cite{hgmd-09,abcg-05,bbgo-05,mf-10,kmt-11,pdc-13,th-15} or  in a quench from the non-interacting model \cite{pce-16, pce2-16}.

In spite of these problems, there are two interesting regimes where the attractive Bose gas can be studied in thermal equilibrium both at zero or finite temperature. The first is the zero density limit (see e.g. \cite{cc-07}), where the system size is sent to infinity, keeping the number of particles finite. The second regime is the one investigated in this work, i.e.  
the infinite system size limit taken with fixed  density of particles but with the attractive interaction sent to zero as system size increases. We will  refer to this as a {\it weakly interacting thermodynamic limit}. 
Importantly, no divergences arise in this regime because an extended gas of attractive bosons is stable for sufficiently small attractive interactions. Furthermore, in this case the system exhibits interesting properties that are absent in the zero density limit such as a quantum phase transition with varying the (rescaled) interaction strength \cite{ccr-00_II, ksu-03,ffp-16}.

Here we compute local correlation functions at zero temperature in the weakly interacting thermodynamic limit. We consider the one-point functions $g_2$ and $g_3$, which are accessible in cold atomic experiments, exploiting their relation to photoassociation and three-body recombination rates \cite{tohp-04,kww-05}. 
Besides the per se interest for a comparison with experimental implementations of the attractive 1D Bose gas \cite{ksfb-02,spth-02}, our results  might be a starting point for the  challenging task of computing correlation functions in arbitrary excited states of the attractive Lieb-Liniger model.

In the first part of this paper we address the exact computation of these correlation functions using the Bethe ansatz method building upon the results of some recent works \cite{pozsgay-11, ffp-16}. In the second part, our findings are compared to the mean-field approach based on the Gross-Pitaevskii equation \cite{dgps-99, pitaevskii}. While for one-dimensional systems it is known that the validity of the latter breaks down for finite interactions \cite{ll-63, gw-00}, it is expected to give accurate results in the small interaction regime. 
Our calculations show explicitly that the results for local correlations obtained by means of the Gross-Pitaevskii equation
become exact in the limit of infinite system size and vanishing interaction, but they are incorrect for finite systems.
This unveils a direct link between the descriptions of the system in terms of the Bethe ansatz method and of the Gross-Pitaevskii equation.

The rest of this manuscript is organized as follows. In section~\ref{sec:model} we briefly introduce the Lieb-Liniger gas and its exact solution. We then discuss in section~\ref{sec:bethe_solution} the weakly interacting thermodynamic limit  and review some recent results \cite{ffp-16} regarding the Bethe ansatz characterization of the ground state in the attractive regime. Section~\ref{sec:correlators} is devoted to the computation of one-point correlation functions $g_2$ and $g_3$, both at finite size and in the infinite system size limit. The Gross-Pitaevskii equation is then introduced in section~\ref{sec:gp}, where we compare the mean-field results to those obtained by means of the Bethe ansatz method. Finally, conclusions are presented in section \ref{sec:conclusions}. Some technical aspects of our work are provided in the appendixes.
\section{The Lieb-Liniger model}\label{sec:model}
We consider the Lieb-Liniger model \cite{ll-63} of $N$ bosons with pointwise interactions on a ring of length $L$ with Hamiltonian  
\be
H=-\frac{\hbar^2}{2m}\sum_{j=1}^{N}\frac{\partial^2}{\partial x_{j}^2}+2c\sum_{j<k}\delta(x_j-x_k).
\label{eq:hamiltonian}
\ee
The interaction strength is related to the one dimensional scattering length $a_{\rm 1D}$ through $c=-\hbar^2/ma_{\rm 1D}$ \cite{olshanii-98} and can be varied via Feshbach resonances \cite{iasm-98} to take either positive or negative values. In the following we set $\hbar=2m=1$ and focus on the attractive regime
\be
c=-\bar{c}<0\,.
\label{attractive_regime}
\ee
The equivalent second quantization Hamiltonian is
\begin{equation}
H=\int_{0}^{L} \hspace{-1mm} \mathrm{d}x  \left\{\partial_x\Psi^{\dagger}(x)\partial_x\Psi(x)+c\Psi^{\dagger}(x)\Psi^{\dagger}(x)\Psi(x)\Psi(x)\right\},
\label{hamiltonian_2}
\end{equation}
where $\Psi^{\dagger}$ and $\Psi$ are bosonic creation and annihilation operators satisfying $[\Psi(x),\Psi^{\dagger}(y)]=\delta(x-y)$.

The Hamiltonian \eqref{eq:hamiltonian} can be diagonalized by means of the Bethe ansatz \cite{ll-63}. 
The $N$-body eigenfunctions are 
\bea
\psi_N\left(x_1,\ldots,x_N\right)&=& \sum_P \prod_{\ell > k } \left[1+ \frac{i \bar c
  ~\text{sgn}(x_\ell - x_k)}{\lambda_{P_\ell} - \lambda_{P_k}}\right]\nonumber\\
&\times & \prod_{j=1}^N e^{i
  \lambda_{P_j} x_j}\,,
\label{wave_function}
\eea
where the sum is over the $N!$ permutations $P$ of the rapidities $\{\lambda_j\}_{j=1}^{N}$. 
The latter are complex numbers which parametrize the different eigenstates of the Hamiltonian, and satisfy the quantization conditions (Bethe equations) 
\be
e^{-i\lambda_jL}=\prod_{k\neq j}^{N}\frac{\lambda_k-\lambda_j-i\bar{c}}{\lambda_k-\lambda_j+i\bar{c}}\ ,\quad j=1,\ldots, N\,.
\label{bethe_eq}
\ee
The momentum ($K$) and energy ($E$) of a given eigenstate are expressed in terms of the rapidities $\lambda_j$ as
\bea
K\left[\{\lambda_j\}_{j=1}^N\right]&=&\sum_{j=1}^N\lambda_j\,,\\
E\left[\{\lambda_j\}_{j=1}^N\right]&=&\sum_{j=1}^{N}\lambda_j^2\,.
\eea

Introducing the density of particles $D$ and the dimensionless interaction $\gamma$ \cite{ll-63}
\be
D=\frac{N}{L}\,,\qquad \gamma=-\frac{\bar{c}}{D}\,,
\label{density}
\ee
the standard thermodynamic limit is defined as $N,L\to\infty$ with $D$, $\gamma$ fixed.
As we already mentioned in the introduction, this is ill-defined in the attractive regime because 
it gives rise to divergences in the ground state energy and local correlation functions \cite{mcguire-64}. 
It is possible to overcome this problem introducing the rescaled interaction 
\be
\kappa=  - \gamma N^2\,,
\label{gamma}
\ee
and defining the weakly interacting thermodynamic limit  as
\bea
N,L\to\infty\,, \qquad D, \kappa\, \ {\rm fixed}\,.
\label{limit}
\eea
As we will show in the following, despite the interaction strength goes to zero as $N$ increases, this limit is non-trivial
and  the physics of the model depends only on $\kappa$.

\section{The ground-state rapidity distribution function}\label{sec:bethe_solution}

In the attractive regime, for any number of particles $N$ the rapidities corresponding to the ground state of the model are always aligned along the imaginary axis and centered around $\lambda=0$ \cite{takahashi,mcguire-64}. In the zero density limit, the rapidities satisfy the well-known string hypothesis \cite{takahashi}, according to which they display a uniform spacing $\bar{c}$ between one another. This is no longer the case in the limit \eqref{limit}, where the rapidities arrange themselves along the imaginary axis according to a non-trivial distribution function. The latter has been recently derived for arbitrary $\kappa$ \cite{ffp-16} as reviewed in this section. We mention that partial results where also presented in previous works \cite{bgm-04,ol-07}, while numerical studies of the ground-state rapidities are reported in \cite{ssac-05,sdd-07}.

\begin{figure}[t]
\includegraphics[scale=0.48]{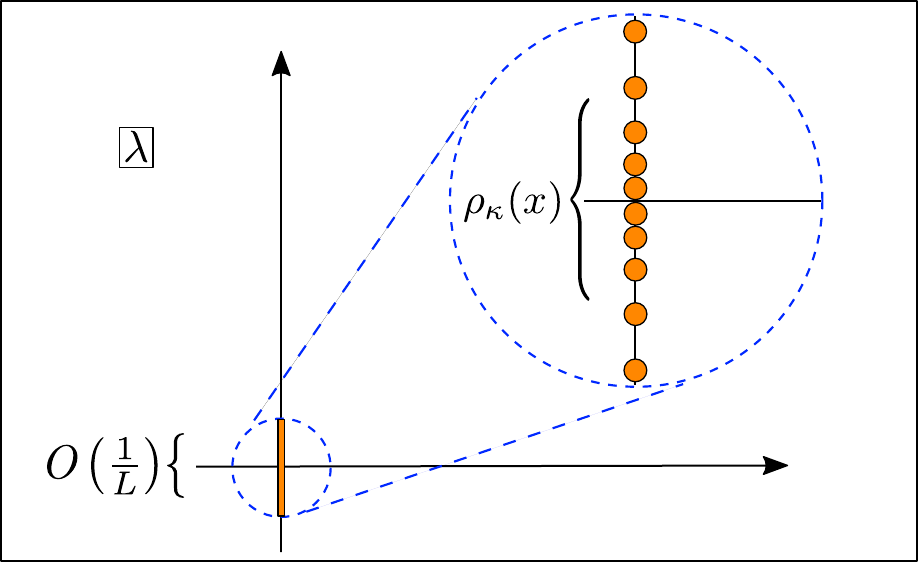}
\caption{Pictorial representation of the ground-state rapidities in the weakly attractive thermodynamic limit.}
\label{fig:1}
\end{figure}

The ground-state rapidities correspond to the unique set (up to permutations) of purely imaginary solutions  $\{\lambda_j\}_{j=1}^N$ of \eqref{bethe_eq}. They are pictorially displayed in Fig.~\ref{fig:1} and in the limit \eqref{limit} they shrink to the point $\lambda=0$. It is then convenient to define the following rescaled ground-state rapidities (which have a different normalisation 
compared to  \cite{ffp-16}):
\be
x_j=-i \lambda_jL\,.
\label{rescaled}
\ee
Plugging \eqref{rescaled} into the Bethe equations \eqref{bethe_eq} and taking the logarithm one obtains the following system of equations for the rescaled rapidities
\be
x_j=\sum_{\substack{l=1\\l\neq j}}^{N}\log\left(\frac{x_j-x_l+\kappa/N}{x_j-x_l-\kappa/N}\right)\,.
\label{eq_log}
\ee

In the limit \eqref{limit} the rescaled rapidities $x_j$ arrange themselves according to a non-trivial distribution function $\rho_{\kappa}(x)$, characterized by the property that for any function $f(x)$ one can write
\be
\sum_{j=1}^{N}f(x_j)= N\int {\rm d}x\rho_{\kappa}(x)f(x)+\mathcal{O}(1)\,.
\label{normalization_1}
\ee 
It has been found \cite{ol-07, ffp-16} that  a critical value $\kappa^*$ of the interaction exists such that $\rho_{\kappa}(x)$ is qualitatively different for $\kappa>\kappa^*$ and $\kappa<\kappa^*$, namely
\be
\kappa^*=\pi^2\,.
\label{critical_k}
\ee 
For $0<\kappa\leq \kappa^*$ the rapidity distribution function is determined as the solution of the integral equation \cite{ffp-16}
\be
x=2\kappa \dashint_{x_{\rm min}}^{x_{\rm max}}\,{\rm d}y\frac{\rho(y)}{x-y}\,.
\label{int_eq_1}
\ee
Here we introduced the principal value integral \cite{pipkin-91}
\bea
\dashint_{x_{\rm min}}^{x_{\rm max}} \frac{f(x)}{x-y}{\rm d}x\equiv  \lim_{\varepsilon\to 0}\Big\{\int_{x_{\rm min}}^{y-\varepsilon}\frac{f(x)}{x-y}{\rm d}x\nonumber\\
+\int_{y+\varepsilon}^{x_{\rm max}}\frac{f(x)}{x-y}{\rm d}x\Big\}\,,
\label{principal_v}
\eea
while $x_{\rm min}$, $x_{\rm max}$ are chosen consistently with the normalization condition 
\be
\int_{x_{\rm min}}^{x_{\rm max}} \rho_{\kappa}(x)=1\,.
\label{normalization_2}
\ee
These equations share some similarities with the large-$N$ limit of the Bethe equations in the Richardson pairing model \cite{dps-04,bgm-04}.

The solution of \eqref{int_eq_1} under the condition \eqref{normalization_2} is \cite{ffp-16}
\be
\rho_{\kappa}(x)=
\left\{\begin{array}{ll}
\frac{1}{\kappa\pi}\sqrt{\kappa-\frac{x^2}{4}},&\quad x\in [-2\sqrt{\kappa},2\sqrt{\kappa}]\,,\\
0&|x|>2\sqrt{\kappa}\,.\end{array}\right.
\label{weak_coupl_rho}
\ee
An important constraint on the ground-state rapidities is $ |x_j-x_k|>{\kappa}/{N}$ \cite{ffp-16}, resulting in the condition
\be
\rho_{\kappa}(x)\leq \frac{1}{\kappa}\,.
\label{condition}
\ee
For $\kappa<\kappa^{\ast}$, $\rho_{\kappa}(x)$ in \eqref{weak_coupl_rho} always satisfies \eqref{condition}. 
The critical point $\kappa^*=\pi^2$ is identified with the value of the interaction such that 
$\rho_{\kappa^*}(x)$ in \eqref{weak_coupl_rho} has a maximum (in $x=0$) equal to $1/\kappa$. 

\begin{figure}[t]
\includegraphics[scale=0.84]{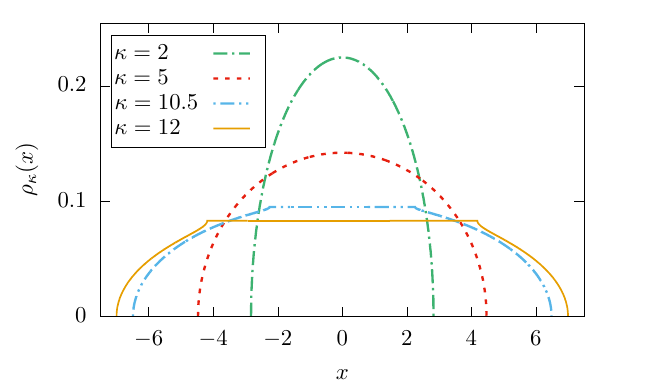}
\caption{Rescaled rapidity distribution $\rho_{\kappa}(x)$ for different values of $\kappa$. 
The values $\kappa=2,5<\kappa^*$ correspond to one quantum phase, while $\kappa=10.5,12>\kappa^*$ correspond to the other.
For $\kappa>\kappa^*$ there is a plateau $\rho_{\kappa}(x)=1/\kappa$ centered around $x=0$.}
\label{fig:2}
\end{figure}

The form of the ground-state rapidity distribution changes qualitatively for $\kappa>\kappa^*$ and it reads \cite{ffp-16} 
\be 
\rho_{k}(x)=
\left\{\begin{array}{cc}
1/\kappa&x\in[-b\kappa,b\kappa]\,,\\
\tilde{\rho}_{\kappa}(x)&x\in[-a\kappa,-b\kappa]\cup[b\kappa,a\kappa]\,,\\
0& |x|>a\kappa\,.
\end{array}\right.
\ee
The parameters $a$ and $b$ are defined as the solution of the non-linear system
\be
\left\{
\begin{array}{ll}
z=b^2/a^2,\\
4K(z)\left[2E(z)-(1-z)K(z)\right]=\kappa\,,\\
a\kappa=4K(z)\,,
\end{array}
\label{ab_system}
\right.
\ee
while the function $\tilde{\rho}_{\kappa}(x)$ is determined by the singular integral equation
\be
 x= 2\log \left(\frac{x+\kappa b}{x-\kappa b}\right)+2\kappa\dashint_{\Omega}\,{\rm d}y\frac{\tilde{\rho}_{\kappa}(y)}{x-y}\,,
\label{singular_eq}
\ee
where the principal value integral is over the domain
\be
\Omega=[-a\kappa,-b\kappa]\cup[b\kappa,a\kappa]\,.
\label{omega_set}
\ee
The solution of \eqref{singular_eq} can be found explicitly to be \cite{dk-93, ffp-16}
\begin{multline}
\tilde{\rho}(x)=\frac{2}{\pi a |x|\kappa^2}\sqrt{(a^2\kappa^2-x^2)(x^2-b^2\kappa^2)}
\\ \times
\Pi\Big(\frac{b^2\kappa^2}{x^2},\frac{b^2}{a^2}\Big)\,.
\label{strong_coupl_rho}
\end{multline}
The functions $K(x)$, $E(x)$ and $\Pi(x,y)$ appearing in \eqref{ab_system}, \eqref{strong_coupl_rho} are the elliptic integrals of the first, second and third kind: 
\bea
K(z)&=&\int_0^{\pi/2}{\rm d}\vartheta \frac{1}{\sqrt{1-z\sin^2\vartheta}}\,,\label{k_function}\\
E(z)&=&\int_0^{\pi/2}{\rm d}\vartheta \sqrt{1-z\sin^2\vartheta}\,,\label{e_function}\\
\Pi(x,y)&=&\int_0^{\pi/2}{\rm d}\vartheta \frac{1}{(1-x\sin^2\vartheta)\sqrt{1-y\sin^2\vartheta}}\,.
\eea
We report in Fig.~\ref{fig:2} the rapidity distribution functions corresponding to different values of $\kappa$ in the two different regimes $\kappa\leq\kappa^*$ and $\kappa>\kappa^*$.
The qualitative difference in the behavior of $\rho_{\kappa}(x)$ for $\kappa\leq\kappa^*$ and $\kappa>\kappa^*$ is a signal of a quantum phase transition. We will return to the nature of the latter in section~\ref{sec:gp}. 
A qualitative change of the distribution of the Bethe rapidities in correspondence of a quantum phase transition 
has been observed also in other integrable models \cite{rfm-12,lm-15}.

Using the above results, the ground state energy per particle can be computed as
\begin{multline}
\epsilon_{\rm gs}(\kappa)=\frac{1}{N}\sum_{j=1}^{N}\lambda_j^2=-\frac{1}{NL^2}\sum_{j=1}^{N}x^2_j
\\=
-\frac{D^2}{N^2}e_0(\kappa)+\mathcal{O}(1/N^3)\,,
\label{energy}
\end{multline}
where
\be
e_0(\kappa)=\int{\rm d}x\rho_{\kappa}(x)x^2\,.
\label{e_0}
\ee
From \eqref{energy} we have that $\epsilon_{\rm gs}(\kappa)\to 0$ as $N\to \infty$ according to the limit \eqref{limit}: hence the ground state energy coincides with that of the non-interacting state. However, we will see that the ground-state local correlation functions in the limit \eqref{limit} are qualitatively different from those of the free case.

\begin{figure}
\includegraphics[scale=0.84]{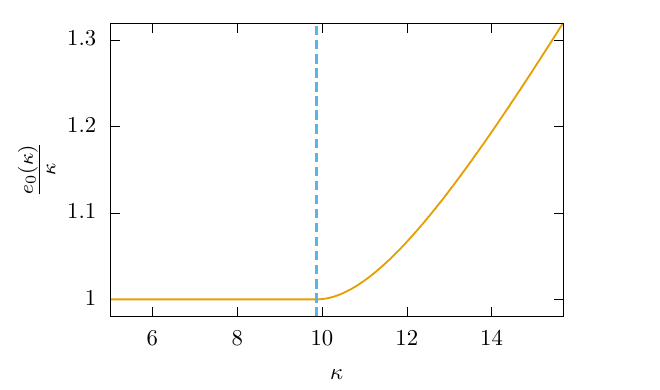}
\caption{Rescaled energy per particle $e_0(\kappa)$ as defined in \eqref{e_0}. The vertical dashed line corresponds to the critical value of the interaction $\kappa^*=\pi^2$, for which $e_0(\kappa)$ exhibits a discontinuity in its second order derivative.}
\label{fig:3}
\end{figure}

It is worth to discuss the relation of the limit \eqref{limit} with other regimes studied in the literature. Consider the large $N$ limit
\be
N\to\infty\,, \qquad L, \kappa\, \ {\rm fixed}\,,
\label{limit_2}
\ee
where $\kappa$ is as usual given by \eqref{gamma}. In this case the value of the density grows indefinitely. 
In this regime, the ground state energy is non-vanishing and given by
\be
\tilde{\epsilon}_{\rm gs}(\kappa)=-\frac{1}{L^2}e_{0}(\kappa)+\mathcal{O}(1/N)\,,
\label{new_energy}
\ee
where $e_0(\kappa)$ is the same as in \eqref{e_0}. The limit \eqref{limit_2} has been studied previously in a number of works \cite{ksu-03,kavoulakis-03,ccr-00_II,kcu-09,kcu-10} and it is known that the system undergoes a quantum phase transition. In particular $e_0(\kappa)$ [as well as $\tilde{\epsilon}_{\rm gs}(\kappa)$, cf. \eqref{new_energy}] exhibits a discontinuity in the second order derivative for $\kappa=\kappa^*$, cf. Fig.~\ref{fig:3}.
Conversely, all the physical quantities depending only on $\kappa$ (such as the local correlations $g_{K}$) 
will have the same value in both limits \eqref{limit} and  \eqref{limit_2}. 

\section{Local correlation functions}\label{sec:correlators}

We address now the computation of the ground-state one-point correlation functions. In particular, we consider the $K$-body functions
\be
g_K=\frac{\langle {\rm GS}|(\Psi^{\dagger}(0))^{K}\Psi^{K}(0)|{\rm GS} \rangle}{D^K}\,,
\ee
where $\Psi$, $\Psi^{\dagger}$ are the bosonic field operators in the second quantization formalism, $D$ is the particle density \eqref{density},
 and $|{\rm GS}\rangle$ the ground state. We focus on the cases $K=2$, $K=3$ which are directly relevant for experimental cold-atomic realizations of bosons confined in one dimension. In particular, $g_2$ is the so called local pair correlation function which can be determined by measures of photoassociation rates \cite{kww-05}. Analogously, $g_3$ is proportional to the three-body recombination rate \cite{tohp-04}. Intuitively, $g_K$ gives information about the probability of finding $K$ bosons in the same position.

\subsection{Finite-size correlators}

The knowledge of the exact (normalized) ground state wave function $\psi_{\rm GS}$ allows the computation of any correlation function. For example, $g_2$ can be expressed as
\begin{multline}
g_2=\frac{N(N-1)}{D^2}\int_0^L dx^{N-2}\psi_{\rm GS}^*(0,0,x_1,\ldots,x_{N-2})\\
\times \psi_{\rm GS}(0,0,x_1,\ldots,x_{N-2})\ .
\label{eq:aux}
\end{multline}
However, the representation \eqref{eq:aux} involves the evaluation of $\sim (N!)^2$ multiple integrals, because of the form of the wave function \eqref{wave_function}. Hence, the r.h.s. of \eqref{eq:aux} can be in practice evaluated only for very small values of $N$.

A remarkable simplification of the problem was obtained  by Bal\'azs Pozsgay \cite{pozsgay-11}, who derived the following alternative representation for $g_K$ by means of algebraic Bethe ansatz methods
\bea
g_K&=&\frac{(K!)^2}{D^K}
\mathop{\sum_{\{\lambda^+\}\cup \{\lambda^-\}}}_{|\{\lambda^+\}|=K} 
\left[\prod_{j>l}\frac{\lambda_j^+-\lambda_l^+}{(\lambda_j^+-\lambda_l^+)^2+c^2)}\right]\nonumber\\
&\times& \frac{\det \mathcal{H} }{\det \mathcal{G}},
\label{eq:start}
\eea
where
\be
 \mathcal{H}_{jl}=
 \left\{
 \begin{array}{cc}
(\lambda_j)^{l-1}
&\text{for}\quad l=1,\dots ,K\,,\\
\mathcal{G}_{jl}&\text{for} \quad l=K+1,\dots , N\,,
\end{array}
\right.
\label{definition_mat}
\ee
and $\mathcal{G}_{jl}$ being the Gaudin matrix
\be
  \mathcal{G}_{jl}=\delta_{jl}\Big(
L+\sum_{r=1}^N \varphi(\lambda_j-\lambda_r)\Big) -\varphi(\lambda_j-\lambda_l)\,,
\ee
with $ \varphi(u)={2c}/{(u^2+c^2)}$.
The sum in \eqref{eq:start} is over all the partitions of the set of rapidities $\{\lambda_j\}_{j=1}^N$ into two disjoint sets $\{\lambda^+_j\}_{j=1}^K$ and $\{\lambda^-_j\}_{j=1}^{N-K}$. Furthermore, the order of the rapidities in both $\mathcal{H}$ and $\mathcal{G}$ in each term of the sum is understood to be given by the ordered set $\{\lambda^+_j\}_{j=1}^K\cup\{\lambda^-_j\}_{j=1}^{N-K}$.

The result~\eqref{eq:start} was obtained in \cite{pozsgay-11}, where only the repulsive regime was considered, but it holds also in the attractive case \eqref{attractive_regime}, because its derivation is purely algebraic. As an additional check, for small $N$ and negative values of the interaction, we numerically verified that \eqref{eq:start} agrees with the result obtained by direct integration of the ground-state wave function \eqref{eq:aux}.

Despite Eq.~\eqref{eq:start} being a great simplification with respect to multiple integral representations of the form \eqref{eq:aux}, it is still not completely satisfying from the computational point of view when large numbers of particles are considered. Furthermore, it is not suitable for the analysis of the thermodynamic limit $N\to \infty$. In fact, it is possible to derive a more efficient representation by direct manipulation of \eqref{eq:start}. This requires a sequence of technical steps which are illustrated in appendix~\ref{sec:app_correlators}, while here we report only the final result.

The results obtained in appendix~\ref{sec:app_correlators} can be written as 
\be
g_2(\kappa,N)=\frac{2}{N\kappa}\sum_{j=1}^{N}\left(x_j^2-x_jw^{(1)}_{j}\right)\,,\label{finite_g2}
\ee
\bea
g_3(\kappa,N)=\frac{1}{N \kappa^2}\sum_{j=1}^N
   \left(3x_j^2w^{(2)}_{j} - 4x_j^3w^{(1)}_{j}+x_j^4\right) \nonumber\\
- \frac{2}{N \kappa}\sum_{j=1}^Nw^{(1)}_{j}x_j + \frac{1}{N^3}\sum_{j=1}^Nx_j\left(w^{(1)}_{j}-x_j\right)\,.\label{finite_g3}
\eea
Here $x_j$ are the rescaled rapidities \eqref{rescaled} while the parameters $w^{(l)}_{j}$ ($j=1,\ldots , N$) are auxiliary variables determined as the solution of the equations
\be
w^{(l)}_{m}+\frac{1}{N}\sum_{j=1}^N\frac{2\kappa\left[w_{m}^{(l)}-w_{j}^{(l)}\right]}{(x_m-x_j)^2-\kappa^2/N^2}=x_m^{l}\,.
\label{eq:discrete_w}
\ee

These formulas allow the exact computation of $g_2$ and $g_3$ for very large number of particles 
(we use up to $N\simeq 2000$ in Sec.~\ref{sec:finite_corrections}). 

\begin{figure*}[t]
\includegraphics[scale=0.89]{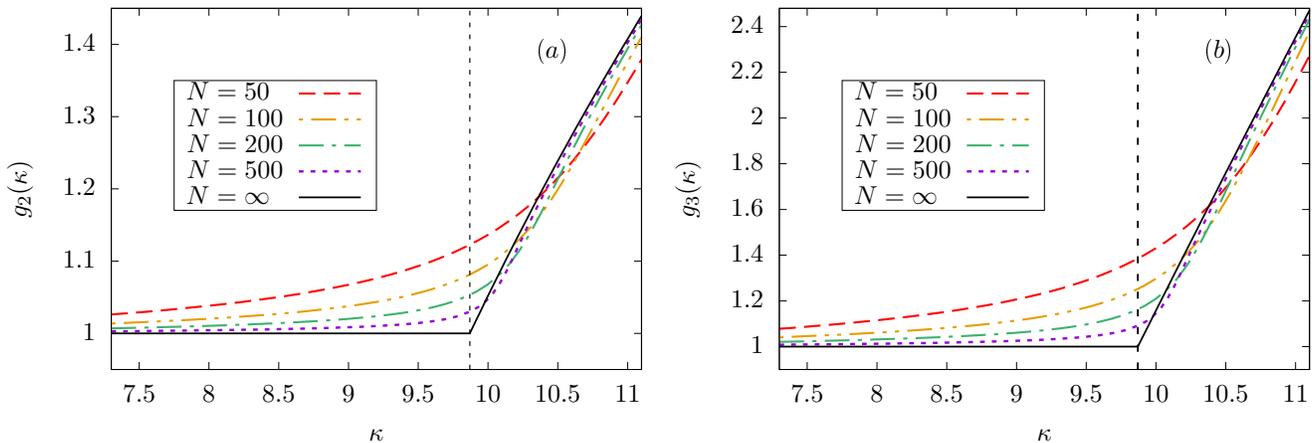}
\caption{One-point correlators $g_2$ and $g_3$ as a function of the interaction $\kappa$ in \eqref{gamma} near the critical point $\kappa^*$. The vertical dashed lines are a guide for the eye corresponding to the critical value $\kappa^*=\pi^2$. The results for increasing numbers of particles $N$ are displayed, showing that large finite-size corrections are observed at the critical point $\kappa^*$.}
\label{fig:4}
\end{figure*}

In Fig.~\ref{fig:4} we report $g_2$ and $g_3$ calculated for several $N$ with this method.
Obviously, no singularity occurs in the behavior of local correlations for finite systems, but 
for $\kappa\sim\kappa^*$ a discontinuity in the first derivatives of both $g_2$ and $g_3$ emerges while
increasing $N$, as we will analytically show in the next subsection.  
Finally, it is worth mentioning that by direct evaluation of \eqref{finite_g2} and \eqref{finite_g3}, one has
\bea
\lim_{\kappa\to 0}g_{2}(\kappa,N)&=&\left(1-\frac{1}{N}\right)\,,\\
\lim_{\kappa\to 0}g_{3}(\kappa,N)&=&\left(1-\frac{1}{N}\right)\left(1-\frac{2}{N}\right)\,,
\eea
namely for $\kappa\to 0$ we recover the ground-state correlators of the free system 
(i.e. the limit $\kappa\to 0$ and the weakly interacting thermodynamic limit commute).

\subsection{Large-$N$ limit}
We now address the computation of the one-point correlation functions in the weakly attractive thermodynamic limit \eqref{limit}. 
Our starting point is given by the formulas \eqref{finite_g2} and \eqref{finite_g3} for finite $N$. 
The calculation is rather cumbersome, but the final results can be easily written down. 
Thus, we anticipate the final results and their discussion, reporting the derivation soon after. 
The full dependence on $\kappa$ of one-point local correlators in the large-$N$ limit is
\bea
g_K(\kappa)&=&\left\{
\begin{array}{ll}
1\,, & 0\leq \kappa \leq \kappa^*\,,\\
g^{s}_K(\kappa)\, & \kappa > \kappa^*\,,
\label{final_result}
\end{array}
\right.
\eea
where $g_K^{s}(\kappa)$ is for $K=2$
\be
g^{s}_{2}(\kappa)=\frac{1}{48}\kappa \left[16 (a^2 + b^2) - (a^2 - b^2)^2 \kappa\right]\,,\label{final_g2}
\ee
and for $K=3$
\begin{multline}
g^{s}_{3}(\kappa)=\frac{1}{240}\kappa^2 \Big[23 a^4 + 82 a^2 b^2 + 23 b^4 \\
- 2 (a^2 - b^2)^2 (a^2 + b^2) \kappa\Big]\,.\label{final_g3}
\end{multline}
The parameters $a$ and $b$ are the solution of the system \eqref{ab_system} and are
easily evaluated numerically for any $\kappa$. Equations \eqref{final_g2} and \eqref{final_g3} give immediately the value of $g_2$ and $g_3$ in the thermodynamic limit. As it is evident from Fig.~\ref{fig:4}, the functions $g_2(\kappa)$ and $g_3(\kappa)$ are not smooth at the critical point $\kappa^*$, where their derivative is discontinuous. For $\kappa<\kappa^*$ the local correlators coincide with those of a non-interacting systems but they rapidly increase for $\kappa>\kappa^*$. This is expected: as the attractive interaction is increased the bosons tend to cluster and have a higher probability of being found in the same position. Note that the opposite behavior is observed for positive values of the coupling $\gamma$ \cite{gs-03,kgds-03,kci-11,pozsgay-11} where the repulsive nature of the interaction is responsible for a decrease in the one-point functions $g_2$ and $g_3$.

\begin{figure}[t]
\includegraphics[scale=0.84]{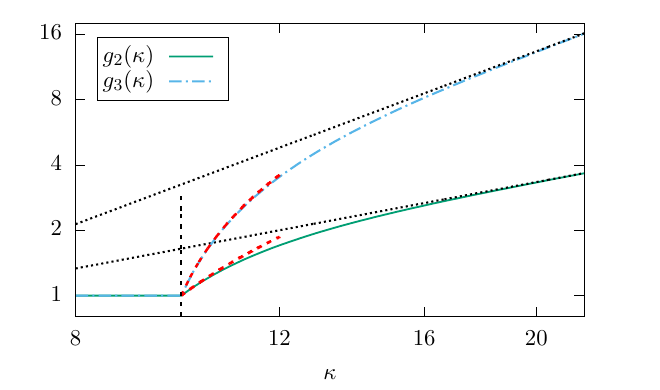}
\caption{One-point correlation functions in the limit \eqref{limit}, as evaluated from \eqref{final_result}. 
Logarithmic scales are used on both axes and the vertical dashed line corresponds to $\kappa^*=\pi^2$. 
Dotted black lines show the asymptotic behavior for large $\kappa$ as given by \eqref{inf_asym2}, \eqref{inf_asym3}, while red dashed lines correspond to the first order expansion in $\kappa-\kappa^*$ as in \eqref{small_g2}, \eqref{small_g3}.}
\label{fig:5}
\end{figure}

Equations \eqref{final_g2} and \eqref{final_g3} also allow for the analysis of the one-point functions in the two limits $\kappa\sim \kappa^*$ and $\kappa\to\infty$. The derivation presents no difficulty and it is sketched in appendix \ref{sec:asymptotics}. Setting
\be
\kappa=\kappa^*+\delta\,,\qquad \kappa>\kappa^*\,,\label{delta}
\ee
in the limit $\delta\to 0^+$ one finds
\bea
g_2(\delta)&=&1+\frac{4\delta}{\pi^2}+\mathcal{O}(\delta^2)\,,\label{small_g2}\\
g_3(\delta)&=&1+\frac{12\delta}{\pi^2}+\mathcal{O}(\delta^2)\,.\label{small_g3}
\eea
Analogously in the limit $\kappa\to \infty$ 
\bea
g_2(\kappa)&=&\frac{\kappa}{6}+\mathcal{O}(1)\,,\label{inf_asym2}\\
g_3(\kappa)&=&\frac{\kappa^2}{30}+\mathcal{O}(\kappa)\,\label{inf_asym3}.
\eea
These asymptotic behaviors are displayed in Fig.~\ref{fig:5}, together with the numerical evaluation of \eqref{final_result}.

\subsubsection{Derivation of the large-$N$ results}

In the large-$N$ limit, the parameters $w^{(l)}_j$ are replaced by a continuous function of the rapidities $w^{(l)}(x)$ such that $w^{(l)}_j\to w^{(l)}(x_j)$. From \eqref{finite_g2}, \eqref{finite_g3} one readily obtains 
\bea
g_2(\kappa)&=&\frac{2}{\kappa}\int {\rm d}x\rho_{\kappa}(x)\left(x^2-xw^{(1)}(x)\right)\,,\label{infinite_g2}\\
g_3(\kappa)&=& \frac{1}{\kappa^2}\int {\rm d}x\rho_{\kappa}(x)\Big(3x^2w^{(2)}(x)-4x^3w^{(1)}(x)\nonumber\\
&+&x^4-2\kappa xw^{(1)}(x)\Big)\,,\label{infinite_g3}
\eea
where the integrals are over the support of the rapidity distribution function $\rho_{\kappa}(x)$. The problem is then reduced to determining the auxiliary functions $w^{(l)}(x)$. 

The idea is to transform the discrete system \eqref{eq:discrete_w} into a linear integral equation for $w^{(l)}(x)$, analogously to what was done in \cite{pozsgay-11} for the repulsive case. Note, however, that in the case considered here one immediately faces the technical issue of dealing with singular integral kernels of the form
\be
\mathcal{K}(x,y)=\frac{1}{(x-y)^2}\,.
\ee
Furthermore, when $x_{m+1}\simeq x_{m}+\kappa/N$ the denominator appearing in the l.h.s. of \eqref{eq:discrete_w} vanishes and near contributions to the sum (corresponding to the terms $|j-m|\ll N$) might be important. The continuum limit of \eqref{eq:discrete_w} is then non-trivial and has to be performed separately for $\kappa<\kappa^*$ and $\kappa>\kappa^*$.

The analysis of the large-$N$ limit of the Bethe equations \eqref{eq_log} (cf. also \cite{ffp-16}), suggests that for $\kappa<\kappa^*$ the near contributions $|j-m|\ll N$ can be neglected in the sum of equation \eqref{eq:discrete_w}; the large-$N$ limit of the latter can then be  cast in the form
\be
w^{(l)}(x)+2\kappa\ddashint\, {\rm d}y\rho_{\kappa}(y)\frac{w^{(l)}(x)-w^{(l)}(y)}{(x-y)^2}=x^{l}\,.
\label{eq:aux_2}
\ee
Here we introduced the Hadamard principal value integral defined as \cite{martin-91,martin-92}
\bea
\ddashint_{x_{\rm min}}^{x_{\rm max}} \frac{f(x)}{(x-y)^2}{\rm d}x\equiv  \lim_{\varepsilon\to 0}\Big\{\int_{x_{\rm min}}^{y-\varepsilon}\frac{f(x)}{(x-y)^2}{\rm d}x\nonumber\\
+\int_{y+\varepsilon}^{x_{\rm max}}\frac{f(x)}{(x-y)^2}{\rm d}x-\frac{2f(y)}{\varepsilon}\Big\}\,.
\eea
Equation \eqref{eq:aux_2} can be explicitly solved for $l=1,2$: one can explicitly verify, making use of \eqref{int_eq_1}, that the following functions are a solution of \eqref{eq:aux_2}
\bea
w^{(1)}(u)&=&\frac{1}{2}u\,,\label{eq:w2}\\
w^{(2)}(u)&=&\frac{1}{3}u^2+\frac{2}{3}\kappa\label{eq:w3}\,.
\eea
Using now the explicit form of $\rho_{\kappa}(x)$ \eqref{weak_coupl_rho} and equations \eqref{infinite_g2}, \eqref{infinite_g3} one obtains
\be
g_{2}(\kappa)=g_{3}(\kappa)=1\,,\qquad 0\leq \kappa\leq \kappa^{*}\,,
\ee
namely for $\kappa<\kappa^{*}$ one-point functions are the same as a non-interacting system.

In the regime $\kappa>\kappa^*$, the computation of the auxiliary functions $w^{(l)}(x)$ is much more involved. 
From section \ref{sec:bethe_solution}, we know that in the interval $(-\kappa b,\kappa b)$ [where $b$ is defined in \eqref{ab_system}] the rescaled rapidities $x_j$ arrange themselves in such a way that for large $N$ they display an equal spacing $\kappa/N$ 
between one another and then one can use the parametrization
\be
x_{j+1}=x_{j}+\frac{\kappa}{N}+\frac{\delta_j}{N}\,,
\ee
where $\delta_j$ vanishes in the thermodynamic limit. Then the corresponding term in the sum  \eqref{eq:discrete_w} apparently diverges as $1/\delta_j$,  but this divergence is canceled if $w^{(l)}_j$ is approximately constant in $(-\kappa b,\kappa b)$\,, namely
\be
w^{(l)}_{j+1}=w^{(l)}_{j}+\frac{\tilde{\delta}_j}{N}\,,
\ee
where $\tilde{\delta_j}$ is also vanishing for $N\to\infty$. Hence, we make the following ansatz for the functions $w^{(l)}(x)$
\be
w^{(l)}(x)= \left\{
 \begin{array}{cc}
C^{(l)}&x\in (-\kappa b,\kappa b),\\
\tilde{w}^{(l)}(x)& x\in \Omega\,,
\end{array}
\right.
\label{ansatz_w}
\ee
where $\Omega$ is defined in \eqref{omega_set} while $C^{(l)}$, $\tilde{w}^{(l)}(x)$ are respectively a constant and a non-trivial function to be determined. This ansatz is well supported by numerical evidence, which provides a posteriori justification for \eqref{ansatz_w}. We now complete the task of explicitly computing the functions $\tilde{w}^{(1)}(x)$, $\tilde{w}^{(2)}(x)$.

First, note that $w^{(1)}(x)$ is  odd with respect to $x=0$. Hence, it has to be $C^{(1)}=0$. Next, we assume that in the region $\Omega$ defined in \eqref{omega_set}, near contributions to the sum in \eqref{eq:discrete_w} can be neglected, so that one can plug the ansatz \eqref{ansatz_w} directly into \eqref{eq:aux_2}. As a result, we find that the function $\tilde{w}(x)$ is determined by 
\bea
&\ &\tilde{w}^{(1)}(x)\left[1+\frac{4\kappa b}{x^2-b^2\kappa^2}\right]\nonumber \\
&\ &+2\kappa\ddashint_{\Omega}\,{\rm d}y\rho_{\kappa}(y)\frac{\tilde{w}^{(1)}(x)-\tilde{w}^{(1)}(y)}{(x-y)^2}=x\,,
\label{eq:temp1}
\eea
for $x\in \Omega$.  Making use of the identity \cite{martin-92}
\be
\ddashint_{\Omega}\,{\rm d}y\frac{\rho_{\kappa}(y)}{(x-y)^2}=-\frac{{\rm d}}{{\rm d} x}\dashint_{\Omega}\,{\rm d}y\frac{\rho_{\kappa}(y)}{(x-y)}\,,
\label{derivative}
\ee
and of \eqref{singular_eq}, Eq.~\eqref{eq:temp1} is easily rewritten as
\bea
2\kappa\ddashint_{\Omega}\,{\rm d}y\rho_{\kappa}(y)\frac{\tilde{w}^{(1)}(y)}{(x-y)^2}=-x\,.
\eea
Rescaling the variables as
\be
\zeta=\frac{y}{a\kappa}\,,\quad \xi=\frac{x}{a\kappa}\,,
\label{rescaling}
\ee
we are left with the simple equation
\be
\left[\ddashint_{-1}^{-r}{\rm d}\zeta+\ddashint_{r}^{1}{\rm d}\zeta\right] \frac{f^{(1)}(\zeta)}{(\zeta-\xi)^2}=-\frac{a^2\kappa}{2}\xi\,,
\label{canonical_form}
\ee
where $r=b/a$ and where we introduced 
\be
f^{(l)}(\zeta)=\rho_{\kappa}(\zeta)\tilde{w}^{(l)}(\zeta)\,.
\label{eq:fw}
\ee
Assuming the continuity of the function $w^{(1)}(x)$, we have that $f^{(1)}(\zeta)$ satisfies the following conditions
\be
f^{(1)}(\pm r)=f^{(1)}(\pm 1)=0\,.
\label{boundary_cond}
\ee
Equation \eqref{canonical_form} belongs to the general family of integral equations with {\rm hypersingular} kernel
\be
\frac{1}{\pi}\left[\ddashint_{-1}^{-r}{\rm d}\zeta+\ddashint_{r}^{1}{\rm d}\zeta\right] \frac{f(\zeta)}{(\zeta-\xi)^2}=\chi(\xi)\,,
\label{eq:family}
\ee
which admits an explicit solution for an arbitrary regular function $\chi(x)$ \cite{db-09}
which is 
\be
f(\zeta)=
\left\{
\begin{array}{ll}
\frac{1}{\pi}\dashint_{-1}^{\zeta}\frac{1}{R(u)}(B+\Phi(u)){\rm d}u\,& \zeta\in (-1,-r)\,,\\ 
\frac{1}{\pi}\dashint_{\zeta}^{1}\frac{1}{R(u)}(B+\Phi(u)){\rm d}u\,& \zeta\in (r,1)\,,\\ 
\end{array}
\right.
\label{general_solution}
\ee
where
\bea
R(u)&=&\left[(1-u^2)(u^2-r^2)\right]^{1/2}\,,\label{r_function}\\
\Phi(u)&=&\dashint_{-1}^{-r}{\rm d}v\frac{\chi(v)R(v)}{u-v}-\dashint_{r}^{1}{\rm d}v\frac{\chi(v)R(v)}{u-v}\,,\label{phi_function}
\eea
and where as usual we used the symbol of dashed integral for the principal value integral \eqref{principal_v}. The constant $B$ is defined as
\be
B=\frac{P}{F}\,,
\label{b_function}
\ee
where
\bea
P&=&\int_{r}^{1}\frac{{\rm d}u}{R(u)}\dashint_{r}^{1}\frac{tR(t)}{u^2-t^2}\left(\chi(t)+\chi(-t)\right){\rm d}t\,,\label{p_function}\\
F&=&\int_{r}^{1}\frac{{\rm d}t}{R(t)}\label{F_function}\,.
\eea
In the special case of \eqref{canonical_form}, from \eqref{b_function} and \eqref{p_function} we have $B=0$, since $\chi(\xi)$ is an odd function. The remaining integrals can be performed analytically and after long but straightforward calculations one obtains
\be
f^{(1)}(\zeta)=\frac{a^2 \kappa  \sqrt{\left(1-\zeta^2\right) \left(\zeta^2-r^2\right)}}{4 \pi }\,,
\label{final_f}
\ee
from which $\tilde{w}^{(1)}(\zeta)$ follows directly from \eqref{eq:fw}. 
One has now all the ingredients to explicitly compute $g_2(\kappa)$ for $\kappa>\kappa^*$. From \eqref{infinite_g2}, using \eqref{strong_coupl_rho} and \eqref{final_f} and after straightforward integration one gets \eqref{final_g2}.

The computation of $\tilde{w}^{(2)}(x)$ can be performed analogously. However, the technical steps are now more involved and its derivation is reported in appendix~\ref{w2_calculation}, together with that of the final result \eqref{final_g3},.

\begin{figure}[t]
\includegraphics[scale=0.84]{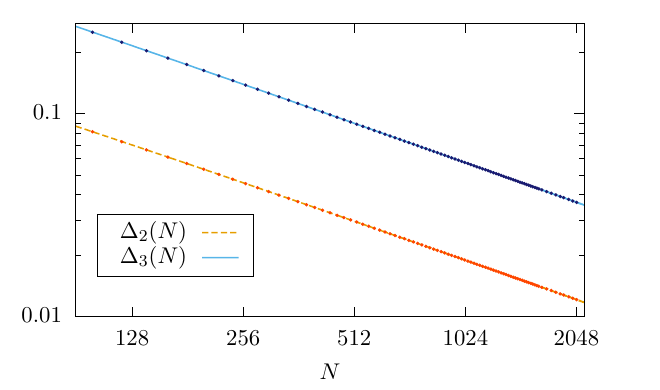}
\caption{Finite size corrections $\Delta_K(N)$ [as defined in \eqref{Delta2}] at the critical point $\kappa^*$ in log scales. 
The dots are the exact numerical values computed using formulas \eqref{finite_g2}, \eqref{finite_g3}, while lines correspond to the fit function \eqref{fit_function}.} 
\label{fig:6}
\end{figure}

\subsection{Finite-size corrections}\label{sec:finite_corrections}

We now investigate the finite size corrections for $g_2$ and $g_3$. 
Away from the critical point, finite size corrections are expected to exhibit an analytical behavior in $1/N$. 
We evaluated numerically the formulas \eqref{finite_g2} and \eqref{finite_g3} for large system sizes up to $N\simeq 1000$
finding that indeed the leading correction is in $1/N$. 
For $\kappa<\kappa^*$, one could even try to tackle this problem analytically, generalizing the techniques of \cite{hlps-05} where 
the Bethe equations in the isotropic spin-$1/2$ Heisenberg chain are studied and the leading corrections  in the system size computed.
Remarkably, the Bethe equations studied in \cite{hlps-05} share a formal analogy with \eqref{int_eq_1}. 
However, the study of one-point functions also requires inspection of finite-size corrections to the auxiliary equation \eqref{eq:aux_2}. 
In any case, these techniques cannot be applied directly at the critical point where a more sophisticated treatment is required. 

At the critical point $\kappa^{*}=\pi^2$, finite-size corrections are more severe as it is clear from Fig. \ref{fig:4}.
To understand their behavior we consider the quantities
\be
\Delta_{K}(N)=g_{K}(\kappa^{*},N)-1\,,
\label{Delta2}
\ee
satisfying $\lim_{N\to\infty}\Delta_K(N)=0$.
For several values of $N$ we computed $\Delta_2(N)$ and $\Delta_3(N)$ from \eqref{finite_g2}, \eqref{finite_g3}, and reported our results in Fig.~\ref{fig:6}. As expected, the dependence on $N$ is not consistent with an analytic behavior in $1/N$. Accordingly, for large $N$ we fit the numerical values of $\Delta_K(N)$ using the function
\be
\ell_K(N)=\frac{A_K}{N^{\alpha_K}}+\frac{B_K}{N}\,.
\label{fit_function}
\ee
For numbers of particles up to $N\simeq 2000$, the best fit for the exponents $\alpha_K$ are 
\bea
\alpha_2 &=&0.667\,,\label{fit:par1}\\
\alpha_3 &=& 0.665\,.\label{fit:par2}
\eea
while the coefficients $A_K$ and $B_K$ are
\bea
A_2&=&2.09\,,\quad B_2=-1.5\,, \label{fit_const1}\\
A_3&=&6.06\,,\quad B_3=-3.1\, \label{fit_const2}.
\eea

The numerical estimates \eqref{fit:par1}, \eqref{fit:par2} suggest the exact value for the exponents to be $2/3$. 
The fitting function \eqref{fit_function} is displayed in Fig.~\ref{fig:6}, showing excellent agreement with the numerical data. In particular, the exponents \eqref{fit:par1}, \eqref{fit:par2} justify the slow approach of $g_K(\kappa^*,N)$ to the asymptotic value $g_K(\kappa^*)=1$ displayed in Fig.~\ref{fig:4}.

\section{The Gross-Pitaevskii equation}\label{sec:gp}

\begin{figure}
\includegraphics[scale=0.84]{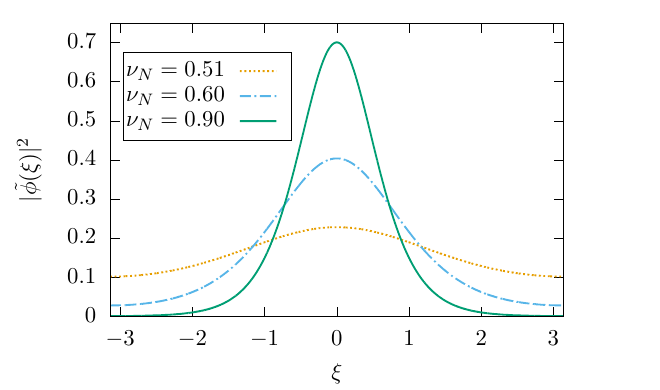}
\caption{Squared absolute value of the ground-state Gross-Pitaevskii wave function \eqref{exact_gs} with $\zeta=0$.}
\label{fig:7}
\end{figure}

In the previous section we considered the computation of one-point functions by means of the Bethe ansatz method. In this section we address the interesting comparison between these exact results and the mean-field approach based on the  Gross-Pitaevskii equation \cite{dgps-99, pitaevskii}. 

While in one dimension the mean-field approximation breaks down for sufficiently strong interaction \cite{ll-63, gw-00}, it is expected to give accurate results in regimes of small coupling \cite{cd-75, yn-77, ll-78,fmt-08, kl-13}. In the case of one-dimensional attractive bosons, this was investigated in \cite{cd-75} in the zero density limit showing that mean-field results for the ground-state energy and reduced one-body density matrix are exact to the leading order in $N$, when $N\to \infty$. It is then of interest to test the mean-field approach also in the weakly attractive thermodynamic limit considered here. This is especially true for the higher-body one-point functions $g_2$, $g_3$ which were not considered in previous studies and for which the question of the accuracy of mean-field calculations is non-trivial.

\begin{figure*}[t]
\includegraphics[scale=0.9]{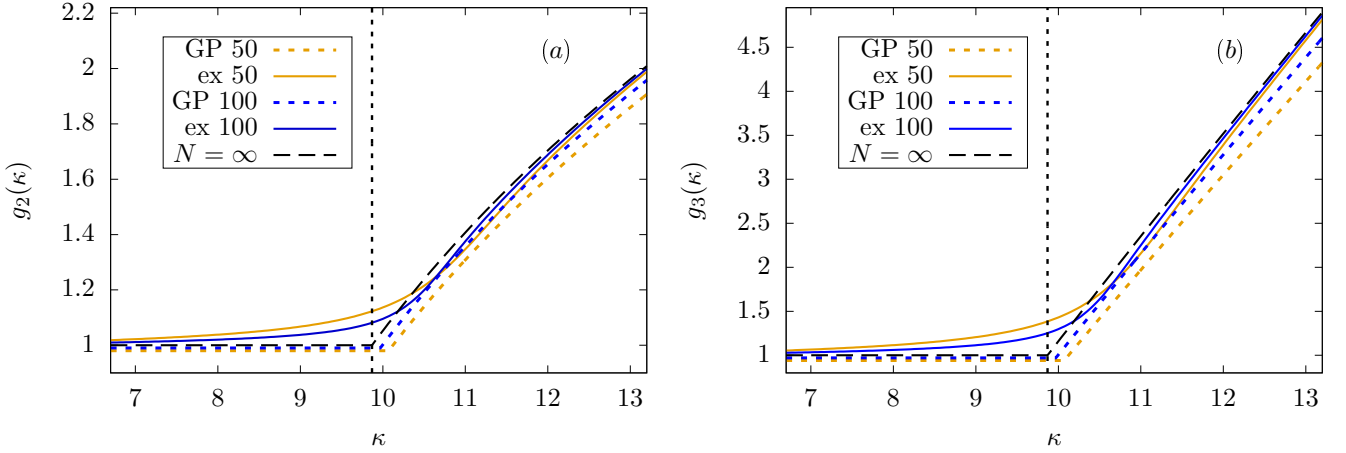}
\caption{One-point correlators $g_2$ and $g_3$ as a function of the interaction $\kappa$ near the critical point $\kappa^*$. The vertical dashed lines  correspond to the critical value $\kappa^*=\pi^2$. The exact numerical values of the correlators (solid lines) are displayed, together with the results obtained by means of the Gross-Pitaevskii equation (dashed lines) which become exact in the limit $N\to\infty$.}
\label{fig:8}
\end{figure*}

In the mean-field description, the ground state is approximated by the product of single-particle wave functions as
\be
\psi_{\rm GS}(x_1,\ldots, x_N)=\prod_{j=1}^{N}\phi(x_j)\,.
\label{factorized}
\ee
The optimal wave function $\phi(x)$ is obtained by minimization of the functional
\be
E[\phi]=\langle \psi_{\rm GS}| H |\psi_{\rm GS}\rangle\,,
\label{functional_energy}
\ee
where $H$ is the Hamiltonian \eqref{eq:hamiltonian}. Following this prescription and using standard techniques, one is directly led to the time-independent Gross-Pitaevskii equation
\be
\left(-\frac{\partial^{2}}{\partial x^2}+2cN\left(1-\frac{1}{N}\right)|\phi(x)|^2\right)\phi(x)=\frac{\mu}{N} \phi(x)\,,
\label{gp_equation}
\ee
where $ 0\leq x\leq L$, and where $\mu$ is the chemical potential, which is introduced to ensure the normalization condition
\be
\int_0^{L}{\rm d}x |\phi(x)|^2=1\,.
\ee
The ground-state wave function then corresponds to the solution of \eqref{gp_equation} with the smallest energy \eqref{functional_energy}. Solutions of \eqref{gp_equation} are known explicitly, and have been studied both in the repulsive \cite{ccr-00} and attractive regime \cite{ccr-00_II}. The exact solution  is written in terms of the rescaled variable
\be
\xi=\frac{2\pi}{L}x \in [0,2\pi],
\ee 
together with the rescaled single-particle wave function
\be
\tilde{\phi}(\xi)=\Big(\frac{L}{2\pi}\Big)^{1/2}\phi(x(\xi))\,.
\label{rescaled_phi}
\ee
Setting $c=-\kappa/(N L)$, it is straightforward to obtain from the previous relations
\be
\Big(-\frac{\partial^{2}}{\partial \xi^2}-2\pi\nu_N(\kappa)|\tilde{\phi}(\xi)|^2\Big)\tilde{\phi}(\xi)=\tilde{\mu}_N \tilde{\phi}(\xi)\,,
\label{gp_equation2}
\ee
where we introduced
\be
\label{nu_parameter}
\nu_N(\kappa)=\frac{\kappa}{2\pi^2}\Big(1-\frac{1}{N}\Big)\,,\qquad 
\tilde{\mu}_N=\frac{\mu}{4\pi^2 D}L\,,
\ee
and where the normalization condition now reads
\be
\int_0^{2\pi}{\rm d}\xi |\tilde{\phi}(\xi)|^2=1\,.
\ee
The solution of minimal energy of \eqref{gp_equation2} under the periodicity condition $\tilde{\phi}(0)=\tilde{\phi}(2\pi)$ can be found in \cite{ksu-03, ccr-00_II}:
\be
\phi_\zeta(\xi)=\left\{
 \begin{array}{ll}
1/\sqrt{2\pi}\,&\  0\leq\nu_N\leq \nu^*,\\
\sqrt{\frac{K(m_N)}{2\pi E(m_N)}}{\rm dn}\left(\frac{K(m_N)}{\pi}(\xi-\zeta)\right)\,& \ \nu_N>\nu^*\,,
\end{array}
\right.
\label{exact_gs}
\ee
where $\nu^*=1/2$, $K(x)$ and $E(x)$ are the complete elliptic integrals in \eqref{k_function}, \eqref{e_function} while ${\rm dn}(x|m)$ is the Jacobian elliptic function \cite{ccr-00}. 
The real parameter $\zeta\in [0,2\pi]$ can be chosen arbitrarily while 
the other real parameter $m_N$ is the solution of the non-linear equation
\be
K(m_N)E(m_N)=\frac{\pi^2\nu_N}{2}\,.
\label{non-linear2}
\ee

We plot in Fig.~\ref{fig:7} the wave function \eqref{exact_gs} for $\zeta=0$ and different values of $\nu_N$. Increasing $\nu_N$ it displays a more distinct peak around $\zeta$, corresponding to the emergence of a bright soliton  \cite{ksu-03, ccr-00_II}. 
Note also that \eqref{exact_gs} apparently breaks translational invariance, but this is not the case because the correct 
ground-state is recovered from \eqref{exact_gs} after averaging with respect to the peak position $\zeta$.
This is the same as the Bethe ansatz wave function  \eqref{wave_function} in which 
(given that the rapidities are purely imaginary) the ground state corresponds to the superposition of a family of many-body wave functions localized around the translated centers of mass of the bosons. 

Exploiting the factorized form \eqref{factorized}, within the mean-field approach one-point functions can be simply obtained from the integral representations such as \eqref{eq:aux}. Averaging with respect to $\zeta$ after performing the integration and expressing everything in terms of the rescaled wave function \eqref{rescaled_phi}, one obtains the mean-field result
\bea
\hspace{-0.5cm}g^{\rm MF}_2&=&2\pi\left(1-\frac{1}{N}\right)\int_0^{2\pi}{\rm d}\xi |\tilde{\phi}_0(\xi)|^4\,,\label{integral_1}\\
\hspace{-0.5cm}g^{\rm MF}_3&=&(2\pi)^2\left(1-\frac{1}{N}\right)\left(1-\frac{2}{N}\right)\int_0^{2\pi}{\rm d}\xi |\tilde{\phi}_0(\xi)|^6\,,\label{integral_2}
\eea
where $\tilde{\phi}_0(\xi)$ is given by \eqref{exact_gs} (with $\zeta=0$).

The integrals \eqref{integral_1}, \eqref{integral_2} can be performed analytically to yield
\bea
g^{\rm MF}_K(\kappa,N)&=&\left\{
\begin{array}{ll}
\tilde{g}^{w}_K\,, & 0\leq \nu_N (\kappa)\leq \nu^*\,,\\
\tilde{g}^{s}_K\, & \nu_N(\kappa) > \nu^*\,,
\label{gp_N_final_result}
\end{array}
\right.
\eea
where
\bea
\tilde{g}^{w}_2&=&1-\frac{1}{N}\,,\nonumber\\
\tilde{g}^{w}_3&=&\left(1-\frac{1}{N}\right)\left(1-\frac{2}{N}\right)\,,
\eea
and
\bea
\tilde{g}^{s}_2&=&\frac{K(m_N)^2(m_N-1)}{3 E(m_N)^2}-\frac{2 (m_N-2)K(m_N)}{3 E(m_N)}\,,\label{gp_g2}
\eea\bea
\tilde{g}^{s}_3&=&-\frac{8 \left(m_N^2-3 m_N+2\right) K(m_N)^3}{30 E(m_N)^3}\,\nonumber\\
&+&\frac{2 \left(8 m_N^2-23 m_N+23\right)K(m_N)^2 }{{30 E(m_N)^2}}.\label{gp_g3}
\eea

These expressions provide the mean-field results for the one-point functions at finite $N$. 
They are displayed in Fig.~\ref{fig:8}, where the comparison with the exact values of the previous section (also reported in the figure) 
show that they are close to the correct results, but show quantitative and qualitative differences. 
First, also for finite $N$ the Gross-Pitaevskii equation predicts a critical point where the derivatives of $g_2$ and $g_3$ are discontinuous, which is of course only an artifact of the mean-field approach.
Second, for $\kappa\leq\kappa^*$ the mean-field finite-size corrections have the opposite sign compared to the exact ones. 
However for $N\to\infty$, the Gross-Pitaevskii equation yields the exact results as we are going to show. 

From \eqref{nu_parameter} we have
\be
\nu_\infty(\kappa)=\frac{\kappa}{2\pi^2}\,,
\ee
so that we simply obtain
\bea
g^{\rm MF}_K(\kappa)&=&\left\{
\begin{array}{ll}
1\,, & 0\leq \kappa \leq \kappa^*\,,\\
\tilde{g}^{s}_K(\kappa)\, & \kappa > \kappa^*\,,
\label{gp_final_result}
\end{array}
\right.
\eea
where $\tilde{g}^{s}_K(\kappa)$ are still given by \eqref{gp_g2}, \eqref{gp_g3}, with the replacement
\be
m_N\to m_{\infty}\,,
\ee
which satisfies
\be
K(m_{\infty})E(m_{\infty})=\frac{\kappa}{4}\,.
\label{non-linear3}
\ee
Remarkably, \eqref{gp_final_result} coincides with the Bethe ansatz result \eqref{final_result}, as it is explicitly shown in appendix \ref{sec:equivalence}.

\section{Conclusions}\label{sec:conclusions}

We  considered ground-state properties of the attractive Lieb-Liniger gas in the limit of large system size and weak 
interactions \eqref{limit}. We addressed the calculations of the physically relevant one-point functions $g_2$ and $g_3$. 
We provided formulas valid at finite size and for arbitrary values of the interaction; furthermore, we showed that in the large-$N$ limit they can be expressed in a simple analytical form. 
We analyzed numerically finite size corrections at the critical point. 
Finally we compared our calculations to the mean-field approach based on the Gross-Pitaevskii equation: 
while the latter provides approximate results for finite systems, it exactly predicts the correct values 
of one-point functions in the large-$N$ limit.
This result was not at all expected: while in the limit \eqref{limit} the interaction $\gamma$  
vanishes with the system size $L$, one would have expected that many-body effects beyond mean-field would contribute to higher-body correlations such as $g_2$ and $g_3$. Our explicit calculations show that this is not the case, providing a direct link between the Bethe ansatz  and the Gross-Pitaevskii equation.

The ground-state calculations performed here might be a useful starting point for the more challenging computation of correlation functions in highly excited states of the model. This would allow us to extend our results to the case of finite temperature and to provide  an essential ingredient to characterize steady states in quantum quenches. 

Having an accurate description of the entire spectrum in the weakly interacting limit might be also useful to 
characterise the height distribution function in Kardar-Parisi-Zhang growth processes \cite{KPZ}.
Indeed, through the replica trick this problem is related to the attractive Lieb-Liniger model \cite{kardareplica}. 
This correspondence has been successfully used to describe the time evolution of the height distribution function in the thermodynamic 
limit for several experimentally relevant situations \cite{droplet,flat,stationary,exp4}. 
Exact calculations for finite systems are still challenging \cite{bd-00}, but the results in this paper represent a 
promising starting point.

\section{Acknowledgments}
We thank Robert Konik, Austen Lamacraft and Andrea Trombettoni for useful discussions surrounding this work. P.C. acknowledges the financial support by the ERC under Starting Grant 279391 EDEQS.

\appendix

\section{Finite-size formulas}\label{sec:app_correlators}

In this appendix we derive the finite size formulas for the one-point correlation functions \eqref{finite_g2}, \eqref{finite_g3} reported in section \ref{sec:correlators}.

We start by rewriting \eqref{eq:start} in terms of the rescaled rapidities \eqref{rescaled}. Exploiting the properties of the determinant we have 
\bea
g_K&=&\frac{(K!)^2}{N^K}
\mathop{\sum_{\{x^+\}\cup \{x^-\}}}_{|\{x^+\}|=K} 
\left[\prod_{j>l}\frac{x_j^+-x_l^+}{(x_j^+-x_l^+)^2-\kappa^2/N^2)}\right]\nonumber\\
&\times& {\rm det}_{N}\left(\mathcal{G}^{-1}\mathcal{H}\right)\,,
\label{eq:start2}
\eea
where the matrices $\mathcal{G}$ and $\mathcal{H}$ are written in terms of the rescaled rapidities as 
\be
 \mathcal{H}_{jl}=
 \left\{
 \begin{array}{cc}
(x_j)^{l-1}
&\text{for}\quad l=1,\dots ,K\,,\\
\mathcal{G}_{jl}&\text{for} \quad l=K+1,\dots , N\,,
\end{array}
\right.
\label{rescaled_mat}
\ee
and 
\bea
  \mathcal{G}_{jl}&=&\delta_{jl}\left(1+\frac{1}{N}\sum_{r=1}^N \frac{2\kappa}{(x_j-x_r)^2-\kappa^2/N^2}\right)\nonumber\\
  &-&\frac{1}{N}\frac{2\kappa}{(x_j-x_l)^2-\kappa^2/N^2}\,.
\eea
The determinant can be rewritten as
\be
{\rm det}_{N}\left(\mathcal{G}^{-1}\mathcal{H}\right)={\rm det}_{K}W\,,
\ee
where
\be
W_{jk}=\left(\mathcal{G}^{-1}\right)_{jm}x^{k-1}_{\sigma(m)}\,.
\label{w_matrix}
\ee
Here, $\sigma$ is the permutation that maps the ordered set $\{x_j\}_{j=1}^N$ into $\{x_{\sigma(j)}\}_{j=1}^N=\{x_j^+\}_{j=1}^K\cup\{x_j^-\}_{j=1}^{N-K}$  Multiplying \eqref{w_matrix} by $\mathcal{G}_{nj}$ and summing over $j$, it is straightforward to show that 
\be
W_{jk}=w^{(k-1)}_{\sigma(j)}\,,
\ee
where $w^{(k)}_j$ are the unique solution of the system \eqref{eq:discrete_w}. Next, we wish to simplify the sum over partitions, analogously to what was done in \cite{pozsgay-11}, where \eqref{eq:start} was studied in the case of repulsive interactions. 

First, we consider the case $K=2$. From \eqref{eq:start2} we have 
\bea
g_2&=&\frac{2}{N^2}\sum_{i,j=1}^N\frac{x_i-x_j}{(x_i-x_j)^2-\kappa^2/N^2}\left(w^{(1)}_{i}-w^{(1)}_{j}\right)\nonumber
\\
&=&\frac{2}{N\kappa}\sum_{j=1}^Nx_j\left(x_j-w^{(1)}_{j}\right)\,,
\eea
where in the first equality we used $w^{(0)}_j=1$ while in last equality we used \eqref{eq:discrete_w}. The computation for $g_3$ is more involved. We define
\bea
\Lambda_{ij}&=&\frac{x_i-x_j}{(x_i-x_j)^2-\kappa^2/N^2}\,,\\
\Gamma_{ij}&=&\frac{1}{(x_i-x_j)^2-\kappa^2/N^2}\,,
\eea
and present the following identity derived in \cite{pozsgay-11}
\bea
\Lambda_{ij}\Lambda_{ik}\Lambda_{jk}=\frac{1}{3}\Big[\Lambda_{ij}\Gamma_{jk}+\Lambda_{jk}\Gamma_{ki}+\Lambda_{ki}\Gamma_{ij}\nonumber\\
-\Lambda_{ji}\Gamma_{ik}-\Lambda_{ik}\Gamma_{kj}-\Lambda_{kj}\Gamma_{ji}\Big]\,.
\eea
Noting now
\bea
{\rm det}_3\left(
\begin{array}{ccc}
1& w^{(1)}_k&w^{(2)}_k\\
1& w^{(1)}_j&w^{(2)}_j\\
1& w^{(1)}_i&w^{(2)}_i
\end{array}\right)=-\left(w^{(1)}_j-w^{(1)}_i\right)\left(w^{(2)}_k-w^{(2)}_j\right)
\nonumber\\
+\left(w^{(1)}_k-w^{(1)}_j\right)\left(w^{(2)}_j-w^{(2)}_i\right)\,,\hspace{1cm}
\eea
and exploiting the properties of the determinant, $g_3$ can be rewritten as
\bea
g_3=\frac{12}{N^3}\sum_{i,j,k=1}^N\Lambda_{ij}\Gamma_{jk}\Big\{-\left(w^{(1)}_j-w^{(1)}_i\right)\left(w^{(2)}_k-w^{(2)}_j\right)
\nonumber\\
+\left(w^{(1)}_k-w^{(1)}_j\right)\left(w^{(2)}_j-w^{(2)}_i\right)\Big\}\,.\hspace{1cm}
\eea
Summing now over the index $k$ and using \eqref{eq:discrete_w} we obtain
\bea
g_3&=&\frac{6}{\kappa N^2}\sum_{i,j=1}^N\frac{x_i-x_j}{(x_i-x_j)^2-\kappa^2/N^2}\Omega_{ij}\,,
\label{g3_intermediate}
\eea
where
\bea
\Omega_{ij}=\Big\{\left(w^{(1)}_j-w^{(1)}_i\right)\left(x^{2}_j-w^{(2)}_j\right)\nonumber\\
-\left(x_j-w^{(1)}_j\right)\left(w^{(2)}_j-w^{(2)}_i\right)\Big\}\,.
\eea
Using \eqref{eq:discrete_w} it easy to show that
\bea
\sum_{j,k=1}^N\frac{x_k}{(x_j-x_k)^2-\kappa^2/N^2}\Omega_{jk}=0\,,
\eea
so that from \eqref{g3_intermediate} we are left to compute
\begin{eqnarray*}
g_3&=&\frac{6}{\kappa N^2}\sum_{j,k=1}^N\frac{x_j\Omega_{jk}}{(x_j-x_k)^2-\kappa^2/N^2}
\nonumber \\
&=&\frac{3}{\kappa N^2} \sum_{j,k=1}^N\Gamma_{jk} \Big\{x_jx_k(x_j-x_k)(w^{(1)}_j-w^{(1)}_k)\nonumber\\
&-&x_jw^{(2)}_j(w^{(1)}_k-w^{(1)}_j)+x_jw^{(1)}_j(w^{(2)}_k-w^{(2)}_j)\nonumber\\
&-&x_kw^{(2)}_k(w^{(1)}_j-w^{(1)}_k)+x_kw^{(1)}_k(w^{(2)}_j-w^{(2)}_k)\Big\}\,,
\end{eqnarray*}
where the last equality comes from symmetrization of the numerator and simple rearrangements. The terms in the last two lines in the above expression can be simplified by summing over $k$ and $j$ respectively and using \eqref{eq:discrete_w}. This yields
\bea
g_3&=&\frac{3}{\kappa N^2} \sum_{j,k=1}^N\Gamma_{jk} \Big\{x_jx_k(x_j-x_k)(w^{(1)}_j-w^{(1)}_k)\Big\}\nonumber\\
&+&\frac{3}{\kappa^2 N}\sum_{j=1}^{N}(x_j^2w^{(2)}_j-x_j^3w^{(1)}_j)\,.
\eea
Finally, we wish to get rid of the double sum. From the simple identities
\bea
 x_jx_k(x_j-x_k)&=&\frac{1}{3}\left(x_k-x_j\right)^3-\frac{x_k^3}{3}+\frac{x_j^3}{3}\,\\
(x_k-x_j)^3&=&(x_k-x_j)\left[(x_j-x_k)^2-\frac{\kappa^2}{N^2}\right]\nonumber\\
&+&\frac{\kappa^2}{N^2}(x_k-x_j)\,,
\eea
and making once again use of \eqref{eq:discrete_w}, we can rewrite
\bea
&\ &\frac{3}{\kappa N^2}\sum_{j,k=1}^N\Gamma_{jk} x_jx_k(x_j-x_k)(w^{(1)}_j-w^{(1)}_k)\nonumber\\
&=&\frac{1}{\kappa N^2} \sum_{j,k=1}^N(w^{(1)}_j-w^{(1)}_k)(x_k-x_j)+\frac{1}{N^3}\sum_{k=1}^N\Big[x_k\nonumber\\
&\times &(w^{(1)}_k-x_k)\Big]+\frac{1}{\kappa^2 N}\sum_{k=1}^Nx_k^3(x_k-w^{(1)}_k)\,.
\eea
As a last step, we exploit the symmetry of the sets $\{w^{(1)}_j\}_{j=1}^N$, $\{x_j\}_{j=1}^N$ to write
\be
\sum_{j,k=1}^N(w^{(1)}_j-w^{(1)}_k)(x_k-x_j)=-2N\sum_{j=1}^Nx_kw^{(1)}_k\,.
\ee
Putting everything together, we finally arrive at \eqref{finite_g3}.
\section{Computation of $w^{(2)}(x)$}\label{w2_calculation}

In this appendix we discuss the derivation of the function $w^{(2)}(x)$ in \eqref{ansatz_w} and the proof of the expression for $g_3$ as given in \eqref{final_g3}. The treatment is analogous to the one presented in the main text for  $w^{(1)}(x)$ and $g_2$, even though technically more involved. 

First, note that in this case the constant $C^{(2)}$ in \eqref{ansatz_w} is non-vanishing and has to be determined independently. Assuming as for $w^{(1)}(x)$ that near contributions to the sum  \eqref{eq:discrete_w} can be neglected in the region $\Omega$ [defined in \eqref{omega_set}], we can plug the ansatz \eqref{ansatz_w} into \eqref{eq:aux_2}, yielding
\bea
\tilde{w}^{(2)}(x)\left[1+\frac{4\kappa b}{x^2-b^2\kappa^2}\right]-C^{(2)}\frac{4\kappa b}{x^2-b^2\kappa^2}\nonumber\\
+2\kappa\ddashint_{\Omega}\,{\rm d}y\rho(y)\frac{\tilde{w}^{(2)}(x)-\tilde{w}^{(2)}(y)}{(x-y)^2}=x^2.
\label{eq:temp2}
\eea
This equation admits a solution for arbitrary $C^{(2)}$. Indeed, consider the translated function
\be
\tilde{\omega}^{(2)}(x)=\tilde{w}^{(2)}(x)-C^{(2)}\,,
\ee
so that
\be
\tilde{f}^{(2)}(x)=\rho_{\kappa}(x)\tilde{\omega}^{(2)}(x)
\ee
satisfies again the condition \eqref{boundary_cond}. Note that $\tilde{f}^{(2)}(x)$ is simply related to $f^{(2)}(x)$ in \eqref{eq:fw} through
\be
\tilde{f}^{(2)}(x)=f^{(2)}(x)-C^{(2)}\rho_{\kappa}(x)\,.
\label{tilde_fw}
\ee
Making use of the identity \eqref{derivative}, Eq.~\eqref{eq:temp2} can be written as
\bea
2\kappa\ddashint_{\Omega}\,{\rm d}y\frac{\tilde{f}^{(2)}(y)}{(x-y)^2}=-x^2+C^{(2)}\,,
\label{eq:aux3}
\eea
and after performing the rescaling \eqref{rescaling}, we simply obtain
\be
\left[\ddashint_{-1}^{-r}{\rm d}\zeta+\ddashint_{r}^{1}{\rm d}\zeta\right] \frac{\tilde{f}^{(2)}(\zeta)}{(\zeta-\xi)^2}=-\frac{a^3\kappa^2}{2}\left(\xi^2-\frac{C^{(2)}}{a^2\kappa^2}\right)\,,
\label{canonical_form2}
\ee
where $r=b/a$ as usual.

This equation is precisely of the form \eqref{eq:family} and thus its general solution is given by \eqref{general_solution}. In order to fix the value of the constant $C^{(2)}$, we observe that \eqref{eq:discrete_w} yields directly the identity 
\be
\frac{1}{N}\sum_{j=1}^{N}w_{j}^{(2)}=\frac{1}{N}\sum_{j=1}^{N}x_j^2\,.
\label{eq:sum}
\ee
Taking the continuum limit we obtain
\be
\int_{-\kappa a}^{\kappa a}\,{\rm d} x\rho(x) w^{(2)}(x)=\int_{-\kappa a}^{\kappa a}\,{\rm d} x\rho(x)x^2\,,
\ee
namely, after simple calculations
\be
\int_{-\kappa a}^{\kappa a}\,{\rm d} x\rho_{\kappa}(x)x^2\,-\int_{\Omega}\,{\rm d} xf^{(2)}(x)-2bC^{(2)}=0\,,
\label{eq:condition}
\ee
where $a$, $b$ are given in \eqref{ab_system}, while $\Omega$ is defined in \eqref{omega_set}.

Equation \eqref{eq:condition} uniquely fixes the value of $C^{(2)}$. In particular, the following prescription can be used to numerically obtaining $C^{(2)}$. One starts with the initial value $C^{(2)}=0$ and considers the corresponding solution of \eqref{canonical_form2} as given explicitly by \eqref{general_solution}. From this, one computes the l.h.s. of \eqref{eq:condition}, which yields a positive real number. Increasing the value of $C^{(2)}$ and repeating these steps, the l.h.s. of \eqref{eq:condition} decreases monotonically: the correct value of $C^{(2)}$ is then simply obtained when the l.h.s. of \eqref{eq:condition} vanishes. Once $C^{(2)}$ and hence $f^{(2)}(x)$ are known, the function $w^{(2)}(x)$ is immediately obtained by \eqref{eq:fw}. Finally, $g_3$ can be numerically obtained from \eqref{infinite_g3} after integration.

In principle, the integrals involved in the exact solution \eqref{general_solution} of \eqref{canonical_form2} can be performed analytically, as it was the case for the function $f^{(1)}(x)$ derived in the main text, cf. \eqref{final_f}. However, now the constant $C^{(2)}$ is non-vanishing and the form of the solution is more involved: the analytical expressions arising in this case are unwieldy, making the full analytical derivation of \eqref{final_g3} extremely tedious. On the other hand, as we described above the numerical value for $C^{(2)}$ and $f^{(2)}(x)$ can be obtained easily and $g_3$ computed accordingly. Then, for arbitrary values of $\kappa$ one can numerically verify that this gives the same value as \eqref{final_g3} to arbitrary numerical precision. The analytical expression \eqref{final_g3}, which is in this way proven numerically, is instead more easily obtained by comparison with the mean-field result, as discussed in appendix~\ref{sec:equivalence}. 

\section{Asymptotics of correlators}\label{sec:asymptotics}
In this appendix we consider the asymptotic behavior of the one-point functions. First, we consider the limit $\delta=\kappa-\kappa^*\to 0^+$. In this limit, $z\to 0$, where $z$ is the solution of the system \eqref{ab_system}. By considering the known series expansions of the functions $K(z)$, $E(z)$, we have from the second equation of the system \eqref{ab_system}
\be
z=\frac{2\delta}{ \pi^2}-\frac{7\delta^2}{4\pi^4}+\mathcal{O}(\delta^3)\,.
\ee
From the third equation of \eqref{ab_system}, using again the series expansion of $K(z)$ we get
\be
a=\frac{2}{\pi}-\frac{\delta}{\pi^3}+\frac{5\delta^2}{4\pi^5}+\mathcal{O}(\delta^3)\,,
\label{a_expansion}
\ee
while from the first equation 
\be
b=\frac{2\sqrt{2}\delta^{1/2}}{\pi^2}-\frac{15\delta^{3/2}}{4\sqrt{2}\pi^4}\,.
\label{b_expansion}
\ee
Plugging \eqref{a_expansion}, \eqref{b_expansion} into \eqref{final_g2} and \eqref{final_g3} we obtain \eqref{small_g2} and \eqref{small_g3}.

Next, we consider the limit $\kappa\to \infty$. In this limit 
$z\to 1$
 in such a way that
\be
(1-z)K(z)\to 0\,.
\ee
Then, from the second equation of the system \eqref{ab_system} and using $E(1)=1$ one obtains
\be
K(z)\sim \frac{\kappa}{8}\,.
\label{temp_2}
\ee
Making use of \eqref{temp_2}, from the first and third equations of \eqref{ab_system} it is then straightforward to obtain
\be
a, b\sim\frac{1}{2}\label{alarge}\,.
\ee
Finally, plugging \eqref{alarge} into \eqref{final_g2}, \eqref{final_g3} we finally arrive at equations \eqref{inf_asym2}, \eqref{inf_asym3}.

\section{Equivalence of infinite-$N$ limits }\label{sec:equivalence}
In this appendix we show the equivalence of the Bethe ansatz and mean-field results \eqref{final_result} and \eqref{gp_final_result} for the one-point functions in the limit \eqref{limit}.

Introducing the parameter
\be
z_{\infty}=\frac{(1-\sqrt{1-m_{\infty}})^2}{(1+\sqrt{1-m_{\infty}})^2}\,,
\ee
one has the following identities \cite{as-64}
\be
K(m_{\infty})=\frac{2}{1+\sqrt{\tilde{m}_{\infty}}}K(z_{\infty})\,,\label{identity1}\\ 
\ee
\bea
E(m_{\infty})&=&(1+\sqrt{\tilde{m}_{\infty}})E(z_{\infty}) \nonumber\\
&-&2\frac{\sqrt{\tilde{m}_{\infty}}}{1+\sqrt{\tilde{m}_{\infty}}}K(z_{\infty})\,, \label{identity2}
\eea
where we defined $\tilde{m}_{\infty}=1-m_{\infty}$.
Using \eqref{identity1}, \eqref{identity2} one can easily show from \eqref{non-linear3} that $z_{\infty}$ satisfies
\be
4K(z_{\infty})\left[2E(z_{\infty})-(1-z_{\infty})K(z_{\infty})\right]=\kappa\,,
\ee
so that from the second equation of \eqref{ab_system} we have $z_{\infty}=z$. Furthermore, exploiting again \eqref{ab_system} one can show
\bea
m_{\infty}&=&\frac{4ab}{(a+b)^2}\,,\label{eq:mab}\\
K(m_{\infty})&=&\left(1 + \frac{b}{a}\right)\frac{a\kappa}{4}\,,\label{eq:kab}\\
E(m_{\infty})&=&(a+b)^{-1}\,.\label{eq:eab}
\eea
Plugging now \eqref{eq:mab}, \eqref{eq:kab} and \eqref{eq:eab} into \eqref{gp_final_result} and after rearrangements one finally obtains \eqref{final_result}.

\end{document}